\documentclass[conference]{IEEEtran}

\usepackage{comment}
\usepackage{epsfig}
\usepackage{subfigure}
\usepackage{amsfonts}
\usepackage{amsmath}
\usepackage{amssymb}
\usepackage{graphicx}
\usepackage{url}
\usepackage{epstopdf}
\usepackage{graphicx}
\usepackage{array}
\graphicspath{ {./images/} }
\usepackage{amsmath,amssymb,amsfonts}
\usepackage{algorithmic}
\usepackage{graphicx}
\usepackage{textcomp}
\usepackage{caption,setspace}

\ifCLASSOPTIONcompsoc
  \usepackage[nocompress]{cite}
\else
  \usepackage{cite}
\fi

\ifCLASSINFOpdf
\else
\fi

\begin{document}
\title{Blockchain Games: A Survey}
\author{\IEEEauthorblockN{
Tian Min, Hanyi Wang, Yaoze Guo and Wei Cai
}
\IEEEauthorblockA{
School of Science and Engineering, The Chinese University of Hong Kong, Shenzhen, China\\
Email: \{tianmin, hanyiwang, yaozeguo\}@link.cuhk.edu.cn, caiwei@cuhk.edu.cn
}}

\IEEEoverridecommandlockouts
\IEEEpubid{\begin{minipage}{\textwidth}\ \\[12pt]
978-1-7281-1884-0/19/\$31.00 \copyright 2019 IEEE
\end{minipage}}

\maketitle

% \author{
%         Tian~Min,
%         Hanyi~Wang,
%         Yaoze~Guo,
%         and~Wei~Cai,~\IEEEmembership{Member,~IEEE}% <-this % stops a space
% \IEEEcompsocitemizethanks{\IEEEcompsocthanksitem T. Min, H. Wang, Y. Guo, and W. Cai are with the School of Science and Engineering, The Chinese University of Hong Kong, Shenzhen, Guangdong
% China, 518172.\protect\\
% % note need leading \protect in front of \\ to get a newline within \thanks as
% % \\ is fragile and will error, could use \hfil\break instead.
% E-mail: caiwei@cuhk.edu.cn
% }% <-this % stops an unwanted space
% %\thanks{Manuscript received April 19, 2005; revised August 26, 2015.}
% }

% \markboth{Journal of \LaTeX\ Class Files,~Vol.~14, No.~8, August~2015}%
% {Shell \MakeLowercase{\textit{et al.}}: Bare Demo of IEEEtran.cls for Computer Society Journals}

% \IEEEtitleabstractindextext{%

\begin{abstract}
With the support of the blockchain systems, the cryptocurrency has changed the world of virtual assets. Digital games, especially those with massive multi-player scenarios, will be significantly impacted by this novel technology. However, there are insufficient academic studies on this topic. In this work, we filled the blank by surveying the state-of-the-art blockchain games. We discuss the blockchain integration for games and then categorize existing blockchain games from the aspects of their genres and technical platforms. Moreover, by analyzing the industrial trend with a statistical approach, we envision the future of blockchain games from technological and commercial perspectives.
\end{abstract}

% Note that keywords are not normally used for peer-review papers.

\maketitle

\section{Introduction}\label{sec:introduction}

\IEEEPARstart{T}{he} blockchain system  \cite{Nofer2017} became one of the most promising technologies since late 2017, nine years after the release of Satoshi Nakamoto's famed Bitcoin whitepaper \cite{Bitcoin}. In addition to the classic application of distributed public ledger, blockchain is now recognized as the foundation of decentralized applications (DApps) \cite{WeiCaiWEHFL2018}. Ethereum \cite{ethereum}, known as the Blockchain 2.0 platform, is designed to facilitate smart contracts \cite{smartcontractinblockchain}, which are open source programs that can be automatically executed without any centralized control. However, the blockchain ecosystem is still struggling with the lack of killer applications. In fact, the initial coin offering (ICO) has been the most public recognized DApp since Ethereum starts its token business. Nevertheless, numerous ICO scams and valueless ``air tokens'' brought a disreputable impression on the emerging technology.

The digital game has the potential to be the rescue of the blockchain. It perfectly fits the nature of the virtual currency ecosystem since it does not have the data entry pollution problems that commonly exist in many other DApps. In fact, the blockchain system fulfills the ultimate dream of many game players: the items they owned in the virtual world are non-fungible, exchangeable, inheritable, and independent to the game service provider. In this paper, we would like to formally define the term of blockchain games: the series of digital games designed and implemented based on the nature of blockchain technologies.

Following are the major benefits blockchain has brought to the game industry. \textbf{Rule Transparency:} Due to the transparent characteristic of blockchain data, players or third-party organizations can audit the smart contract based games rules, which was hidden in the centralized server in traditional games. The transparent game rules will enhance the trustworthy of the game operation. \textbf{Asset Ownership:} In traditional online games, the virtual properties, including credits, items, and avatars, belong to game operators since all data are stored in the game operators' server. In contrast, the blockchain game player owns their game assets, because all virtual assets will be bound to the players' own address, which allows the players to control and manage everything. The ownership enables the game assets to be independent of specific game operators, which allow the players to retain their digital properties and in-game relationships, even after the game stops its operation. Thanks to the ownership feature, virtual assets in the blockchain ecosystem have tremendous market liquidity. The possibility of asset trades across different games and blockchain platforms stimulates the players to better engage in the game economy. Moreover, the liquidity enables potential new profit model for game operators: the value of the assets will be increased if the demand surpasses the supply, which should be restricted by immutable smart contracts. The game operators may benefit from the value increase of the tokens they issued. \textbf{Assets Reusability:} Intrinsically, the blockchain is an open database that hosts data and executable programs. Game developers can leverage blockchain to design an ecosystem that allows players to reuse their characters and virtual items across different games. To this end, newly launched games can directly inherit game assets from the existing ones. \textbf{User-Generated Content (UGC):} UGC in traditional games are restricted in the specific game, thus, belongs to the game operator. In contrast, these contents can be preserved by the players, thus, has the potential to be shared among multiple games. This benefit will, in turn, encourage the players to participate in the construction of new content.

According to the above advantages, the industry has started its venture on integrating blockchain into gaming systems. The Blockchain Game Alliance\footnote{http://blockchaingamealliance.org/} was formed in September 2018 to explore alternative uses of blockchains in video games. In contrast, there is still a blank in academia to study the impact of blockchain on the future form of games. In this work, we survey the state-of-the-art blockchain games and analyze their trend from a statistic approach. The remainder of this paper is organized as follows. We briefly introduce the blockchain technology in Section~\ref{sec:blockchain} and illustrate the system architecture for blockchain game in Section~\ref{sec:architecture} to provide an overview. Afterward, We survey the state-of-the-art blockchain games by categories in Section~\ref{sec:category}. We also perform data analysis to summarize and predict the future trend of blockchain games in Section~\ref{sec:analysis}.  Section~\ref{sec:conclusion} concludes the article and envision the future of blockchain games.
% Challenges and opportunities for future blockchain systems are discussed in Section~\ref{sec:challenges}.

\section{Background: Blockchain}\label{sec:blockchain}

By definition, a blockchain is a continuously growing chain of data blocks, each of which contains a cryptographic hash of the previous block, a time-stamp, and its conveyed data \cite{Nofer2017}. Due to the existence of the cryptographic hash, the data stored in a blockchain are inherently resistant to modification: if one block of data is modified, all blocks afterward should be regenerated with new hash values. This feature of immutability is fundamental to blockchain applications. The blockchain data structure, together with the peer-to-peer (P2P) system and the proof-of-work (PoW) \cite{Hashcash2002} consensus model, makes the decentralized ledger for cryptocurrencies became a reality \cite{Bitcoin}. On this basis, smart contracts, the programs hosted and executed by the blockchain platforms, process business logic in a transparent and autonomous way. By leveraging smart contract, blockchain games enable data storage and key functionality execution while avoiding the services of a game operator.

% In this work, we define write smart contracts to develop a decentralized toll collection system for edge service sharing among multiple parties, which perfectly demonstrate the benefits of DApps.

\section{Blockchain Game Architecture}\label{sec:architecture}

\begin{figure}[htp]
\centering
\includegraphics[width=8.8cm]{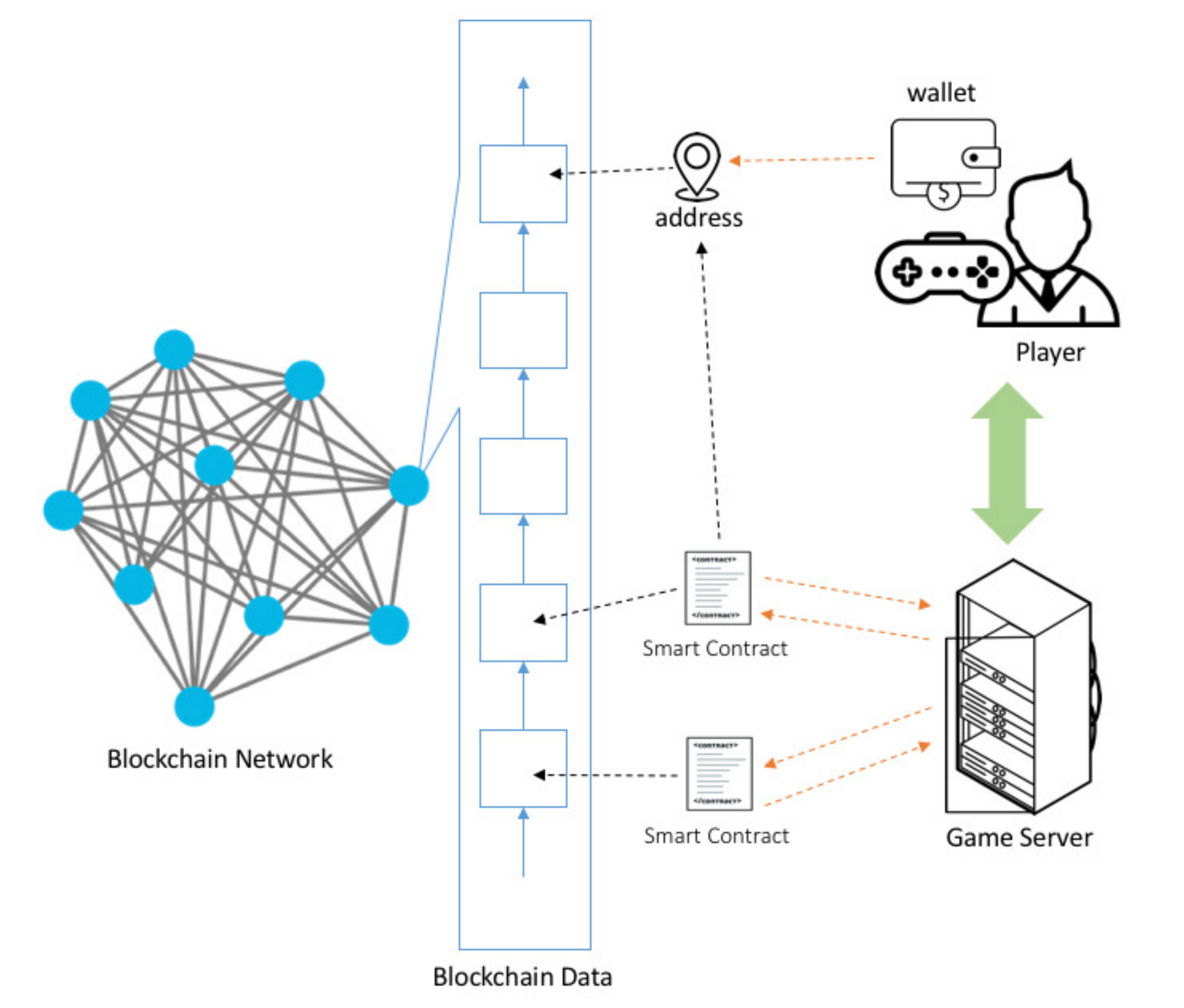}
\caption{Architecture for Blockchain Games}\label{fig:architecture}
\end{figure}

Fig. \ref{fig:architecture} illustrates the architecture for conventional blockchain games. Different from traditional games, the blockchain game players should register an address in the corresponding blockchain before they start their gaming sessions. This blockchain address, accessed by a wallet program later, will serve as the destination of virtual assets for its corresponding player. On the other hand, the game server should offload some core functions, e.g. the ones manipulating the players' virtual assets, to the blockchain as smart contracts. These smart contracts are open source programs written in the blockchain.

From the perspective of the blockchain, to support digital games implies high-performance requirements, including transactions per second (TPS), response latency, etc. Conventional blockchain adopting Proof-of-Work (PoW) \cite{Hashcash2002} consensus models, e.g. Ethereum, failed to satisfy these needs. Hence, some new consensus models have been proposed by novel blockchain platforms, including EOS\footnote{https://eos.io/}, Tron\footnote{https://tron.network/}, Neo\footnote{https://neo.org/}, Qtum\footnote{https://qtum.org/zh},  Nebulas\footnote{https://nebulas.io/index.html,}, etc. In this paper, we mainly focus on Ethereum and EOS, since they have the largest market cap among all.

\section{Category of Blockchain Games}\label{sec:category}
To illustrate an overview of commercial blockchain games, we briefly introduce 23 representative games from various platforms in this section. These games are classified into 4 categories by the major benefits brought by blockchain.

\subsection{Rule Transparency}

The rule is the most critical element of a game since all players are making their choices based on the preset rules. However, not all games rules are transparent to the players, since they are always hidden in the centralized game server. For example, the production of random numbers, commonly used in determining the delivery of virtual items through a lottery, is usually unsupervised. The game developers tend to claim a higher winning probability in order to attract players, while the players can not audit the process. This issue became more critical when the value of virtual items increased. In blockchain games, critical game rules can be implemented as smart contracts to guarantee their transparency. In this way, the players can supervise the lottery smart contracts and prevent developers from arbitrarily altering lottery functions. Apparently, the transparent characteristic of blockchain improve the trustworthiness of the games, especially for those ones requires high reliability, e.g., online casinos.

% \begin{itemize}
\textbf{Satoshi Dice\footnote{https://satoshidice.com/en}; Bitcoin; 4/2012}: Launched in 2012 as the very first gambling game on the blockchain, Satoshi Dice contributed more than half of all the Bitcoin transactions. The rules are simple: first, players send Bitcoins to the jackpot, a certain neutral address on the blockchain, to place their bets on a number. Second, the system will randomly choose a lucky number. If the chosen number is smaller than the lucky number, players win, otherwise, they lose. Although the game mechanism is simple, it's a milestone to blockchain gamification.

\textbf{PoWH 3D (Proof of Weak Hands 3D)\footnote{https://powh.io}; Ethereum; 2/2018}: A Ponzi Scheme running on Ethereum. Players use ETH to buy tokens, the first token costs 0.0000001ETH. The price will be 0.0000001ETH higher when another token is bought. If a token is sold, the price will be 0.0000001ETH lower. Each deal will charge a 10\% commission fee for both the buyer and the seller. This fee will be averagely distributed among the token holders. Smarts contract can prevent Ponzi bankers from escaping with the money. PoWH 3D triggered a Ponzi heat on Ethereum. Many similar games flow into the market, such as PoWL, PoWC, PoWD and so on. It's called a ``Clone War''\footnote{https://coincryptonews.com/tag/p3d-clones/} of P3D games.

\textbf{FOMO 3D\footnote{https://exitscam.me/play}; Ethereum; 7/2018}: It is one of the most famous blockchain Ponzi Schemes, with the highest jackpot of 3 million dollars. The game has a 24 hours countdown. Once a player buys a new key, it will add 30 seconds. Meanwhile, the cost of the keys will get higher. When the countdown touches 0, which means there is no more new key buyer, the last buyer of the key will get 48\% of the jackpot. Rest of the jackpot will become a bonus that distributes to all shareholders with a proportion linked with the number of the keys they hold. It's one of the most remarkable blockchain game on Ethereum.
%Airdrop: 1\% of all game revenue will go into the airdrop pool. If players spend more than 0.1 ETH to buy keys, they will have chances to get ''airdrop''. P3D: P3D is the token issued by the game developers. P3D holders can participate in a bonus distribution. Choose a team: When players buy keys, they will also choose a team from ''sneak'', ''bull'', ''whale'' and ''bear'', which will decide the proportion of bonus distribution.

\textbf{BetDice\footnote{https://www.betdice.one/}; EOS; 9/2018}: The player will roll a dice: if the selected number is smaller than the result, the player wins, otherwise loses. The target number is chosen between 1 to 100 by a smart contract to ensure fairness. BetDice is one of the most popular DApps on EOS at the end of 2018. Because it's run on the EOS chain, there is no gas fee or time delay, which attracted much more players than casinos on Ethereum.

\textbf{FarmEOS\footnote{https://www.farmeos.io/}; EOS; 10/2018}: FarmEOS is a virtual casino that combines various gambling games like Baccarat, Keno and Roulette together. It has a shareholder system to offer a dividend. By the end of 2018, FarmEOS had 29,000 weekly active users and transaction volume of 6.6 million EOS. In the meantime, EOS-based gambling games experienced explosive growth: 147 out of 240 games on EOS are gambling games\footnote{https://DAppradar.com/rankings/protocol/eos/category/gambling}. These highly homogeneous games contributed over 90\% of transaction volume.

%However, gambling games on EOS will be close to saturated one day, and fierce market competition will be unavoidable at that time.

\textbf{TRONBet\footnote{https://www.tronbet.io/}; TRON; 10/2018}: TRONBet is a gambling game based on TRON blockchain platform where the players can gamble with TRX, the token of TRON. The game has 3 modes, Roll, Moon and Ring. In the Roll mode, the players can adjust the slide to change their prediction of the lucky number. In the Moon mode, the player needs to escape before time elapses to get a reward. In the Ring mode, the players need to guess the color of a random wheel. Another dividend mechanism is that once the players bet, they can get ANTE, the token of Tronbet. They can freeze ANTE to receive dividends from Tronbet. ANTE can also be transacted in exchanges.

\subsection{Asset Ownership}
User agreements in the conventional games usually state the players' right to use the game account, but lack of emphasis on the ownership of the virtual properties. In contrast, blockchain game players have more genuine ownership of their virtual properties. Blockchain games allow players to trade their game properties into cryptocurrencies that can be circulated around the internet. What's more, the game items will not only be valuable in a particular game.

\textbf{CryptoKitties\footnote{https://www.cryptokitties.co/}; Ethereum; 11/2017}: CryptoKitties is a simulation game about breeding, collecting and trading kitties, which are actually indivisible and unique ERC-721 tokens. The operations on kitties like selling and breeding are functions implemented in the smart contracts. The game has a web application interface to help players better interact with Ethereum. As one of the most famous blockchain games launched in November 2017, it has caused a great sensation: the rarest kitties worth more than one million dollars. It created a record of 40,000 active users in one day, and once jammed the Ethereum network. Till the second year in April, the total turnover reached 43,067.04 ETH, which is about 200 million dollars. However, it has many drawbacks. The most criticized point of the game is the poor game experience. Many people consider it as a blockchain fund plate rather than a game. In addition, its playability is limited: even Tetris has more fun then CryptoKitties. According to the recent statistics from DAppRadar\footnote{www.DAppradar.com}, the transaction volume has dropped to only 1\% of the peak value, only 4,000 of 720,000 rare kitties were sold, and less than 1,000 players remain active.

\textbf{Etheremon\footnote{https://www.etheremon.com/}; Ethereum; 12/2017}: Etheremon is a blockchain edition of Pokemon, which overcame CryptoKitties in terms of playability. Players can capture, train Etheremon, and participate in combat with each other. Similar to CryptoKitties, every Etheremon is a unique token. Its developers focus on balancing the game and improving the gameplay. Many new models like monster types and skill system were added in order to make the game more interesting. Although the extra parts may not be completely implemented on smart contracts, these efforts made Etheremon ranked at the top 5 Ethereum games for a long period.

\textbf{EtherGoo\footnote{http://ethergoo.io/}; Ethereum; 4/2018}: After the success of ``Hot Potato'' Mode and pet cultivation games, simulation and strategy games like EtherGoo emerged. At the beginning of the game, players own a factory and a cat, which will earn you resources: Goo. Players use Goos to upgrade their factories or buy different types of cats. A reward will be given to the most productive player every day. What's more, players can send their cats to attack other factories in order to plunder resources. In the early May of 2018, the game successfully reached the top three blockchain games with the largest transaction volume: more than 200 ETH transactions in 24 hours. Due to its popularity and the big success, \textbf{TronGoo\footnote{https://trongoo.io/}} has been launched on TRON blockchain platform in January 2019.

\textbf{Gods Unchained\footnote{https://godsunchained.com}; Ethereum; 7/2018}: Gods Unchained is a blockchain card game, who is similar to the traditional TCG (Trading Cards Game) like Hearthstone\footnote{https://playhearthstone.com}. In traditional TCGs, players purchase cards from the game operators, who can infinitely issue cards. In Gods unchained, each set of cards are issued by smart contracts in limited quantity. Therefore, the players can only trade cards with others if the packages have sold out. Gods Unchained is one of the most successful blockchain game in 2018 summer. It has exquisite characters and fancy animations, which helped it earned more than 1.3 million dollars in half a month, according to the announcement from its official website. However, Gods Unchained is unfriendly to the players who cannot spend much money on games because of the high price of its cards.

\textbf{0xUniverse\footnote{https://0xuniverse.com}; Ethereum; 7/2018}:  0xUniverse creates a virtual universe where players can build spaceships, explore the galaxy, and colonize planets, which are all digital collectibles. The discoverers will extract resources and carry out research that allows them to conquer the remotest corners of the galaxy. In addition, players can jointly contribute to the story and unravel the mystery of the universe.

\textbf{EOS Knights\footnote{https://eosknights.io}; EOS; 8/2018}: Developed by a Korean team, EOS Knights is the first mobile game that runs on EOS. Players can fight with Goblins, the monsters in the game, using their own characters. If the characters die, the game will settle items players earned during this expedition. Players can trade, craft, and equip more than 55 kinds of items and weapons. The more weapons equipped on the characters, the more powerful they will be, and the more reward players will get after each game. Compared to previous blockchain games, EOS Knight's user interface is much more gorgeous and the interactive latency is much smaller thank to the Delegated Proof-of-Stake consensus model in the EOS platform. EOS Knight attracted over 500 players once it launched and the number has increased to more than 5,300 up to the end of 2018\footnote{https://DAppradar.com/eos/652/eos-knights}, which make it one of the most popular blockchain games.

%It once again proves that what attracts players is not the title of ``blockchain'', but the gameplay itself.

% \textbf{My Crypto Heroes\footnote{https://www.mycryptoheroes.net/}; Ethereum; 9/2018}: My Crypto Heroes is a multi-player role-playing game (RPG). Players can level up their heroes through quests and the player vs player battles. Heroes and items are secured by the blockchain as ERC-721 token, their value grows as players progress through the game. My Crypto Heroes occupied the second place of all the blockchain games at the end of 2018.

% \textbf{Sand Hero\footnote{https://sandhero.io/}; EOS; 11/2018}: Sand Hero is an RPG similar to EOS Knight. Players explore a dungeon and gain trophies from each expedition. There are various items and a trading system. According to the author, Sand Hero is completely run on EOS and 99\% of the data are on the chain. It ranked top 3 EOS games by the end of 2018. The author claimed that there will be further upgraded for Sand Hero in order to improve the gameplay. The emergence of the independent game on the public chain like EOS is a remarkable event, which may drive further development on the blockchain game industry.

\textbf{EOSDOTA\footnote{https://www.eosdota.com/}; EOS; 1/2019}: EOSDOTA is a PVP trading card game. Players will select three heroes to form a team. Each hero has its own rarity, type and combat power. In every single round of a game, players need to designate one hero to battle. The damage caused by a hero is randomly chosen between 1 and its combat power. Players can gain hero cards in three ways: 1. directly buy cards from the developers; 2. attend championship contest; 3. win cards from other players in plunder mode.

\textbf{HyperSnakes\footnote{https://www.hypersnakes.io/e/d/d/index.html}; Ethereum \& TRON; 5/2019}: HyperSnakes is the blockchain version of the famous multi-player online game Slither\footnote{Slither.io}. Players control a snake which can grow longer by collecting dots or eating smaller snakes in the arena. To make the game more interesting, developers are going to add a number of features such as shields, speed boosts and more. These power-ups can greatly help a player to survive in the arena. HyperSnakes adopts a free-to-play model and pay-to-play rooms where players are able to compete with each other and earn rewards in cryptocurrency for their performance in the game.

\subsection{Asset Reusability}

In contrary to the centralized database server of traditional games, the blockchain is a more open platform for game developers. The players' data can be accessed by smart contracts or a set of application programming interfaces (APIs). Although there is still tons of work to be done, including setting up a stable economic system and consistent numerical design for multiple games, it is exciting to imagine if a player can use his/her own character in RuneScape\footnote{https://www.runescape.com/splash} to continue adventures in World of Warcraft\footnote{https://worldofwarcraft.com/en-gb/}. Brought by CryptoKitties, KittyVerse is a game asset reuse project aiming to build an ecosystem spreading kitties in more blockchain games. The project has released a license called Nifty\footnote{https://www.niftylicense.org/} for developers who are interested in reusing the CryptoKitties assets. Followings are some of them.

%https://coiniq.com/cryptokitties-guide/#CryptoPurr

\textbf{KittyRace\footnote{https://kittyrace.com/}; Ethereum; 3/2018}: KittyRace is a blockchain game that reuses the assets from CryptoKitties. It allows CryptoKitty owners to race their kitties against others for Ethers. According to the whitepaper, a kitty's hustle is determined by three digital dice rolls: a provably fair random roll, genetic makeup, and position. The first is purely random, the second is determined by specific attributes of the players' kitties, and the third is affected by the order in which the player joined the race (the first two kitties to join get a slight advantage). The race will be started by calling the KittyRace smart contract after the players pay their entry fee. The player who executes the contract will be reimbursed in the form of 10\% of the initial fee.

\textbf{KittyBattle\footnote{https://alpha.kittybattles.io/}; Ethereum; 12/2018}: KittyBattles is a kitty combat game that lets players' form their kitties into a team and battle other teams with water balloons, pillows and other deadly items. According to their blog, KittyBattles conducts the battles off-chain to reduce the overhead of Ethereum smart contract invocations. While KittyBattles is intended as a turn-based player-vs-player game, the game is still in alpha and we had a hard time finding real people to battle with.

\textbf{KotoWars\footnote{https://kotowars.com/}; Ethereum; 12/2018}: KotoWars invites the players to fill decks with their cryptokitties and then try to reduce opponents health points to zero. Each CryptoKitty is assigned a defense score, an attack score and an element. The first two parameters depend on the cat's attributes at CryptoKitties, and the elements will introduce a magical system that diversifies the gameplay. One game session requires at least 32 CryptoKitties to be started, but if the player doesn’t have enough they can use computer-generated cards instead. Another interesting feature is that the KotoWars ties into KittyHats\footnote{https://kittyhats.co/}, another KittyVerse-based game so that the customized hats that the players bought in KittyHats will appear in the game.

\subsection{User-Generated Content}

UGC is any form of content that has been posted by users rather than the developers/operators on online platforms. UGC in games keeps updating the game elements, which is critical in retaining veteran players. In contrast to traditional game communities like Steam Workshop\footnote{https://steamcommunity.com/workshop/}, storing UGC in blockchain emphasizes the ownership of the created item and the intellectual property of the author. Thus, it helps remove commercial intermediary to provide an open source community atmosphere.

%Since all the game data are stored on the chains, blockchain games are capable of letting players edit or contribute to game content. Blockchain system acts like a cloud server that can receive information from players, while players acquire game data from the blockchain. Thus, every player can become part of the game. This is huge progress on improving blockchain games' gameplay. Mod is an important part of game developing.

\textbf{Cell Evolution\footnote{http://www.cellevo.net/}; Nebulas; 5/2018}: It is a unique game that uses blockchain to form a player community. Each player in this game act as an individual in a huge population of cells. In this population, cells need to balance several attributes to survive. In the game, players decide their evolution strategies through four buttons: reproduction, evolution, mutation, and dormancy. Each click on buttons is one round of a game. Players need to explore the evolution and mutation pattern to gain a higher score by surviving longer and reproduce more cells. As the round number increase, the game will get harder. Players can update their DNA information into ``the worlds'' when his game is over. This is actually uploading data to the blockchain. There are 8 worlds in this game. Every players' DNA will affect the survival of the entire population. Only balancing the attributes can survive. The attributes' balance check is defined as follows: $tolerance balance check=(a + s + r)/3$, where $a$ is adaptivity, $s$ is survivability and $r$ is reproductivity. For each component in the balance check, it should fall between 0.5 and 1.5 of balance check, otherwise, the world will collapse.

\textbf{CardMaker\footnote{https://www.cardmaker.io/}; Nebulas; 7/2018}: CardMaker is a UGC game that ``focuses on content design, battles, and trading.'' The players are alchemists who use their own cards to participate in turn-based combats. In CardMaker, the game developer only creates game frameworks and basic rules, while the players are encouraged to design their own cards, characters, levels, and trade them to gain the upper hand in battles and achieve their goals in the game. Under the token economy, the players can vote all the design, the content, the update and the designer if they like. Also, the players can share and trade their self-designed cards with other players.

\textbf{Last Trip\footnote{http://lasttrip.matrixDApp.com}; Ethereum \& TRON \& Nebulas; 10/2018}: Last Trip is a storybook in which the players will experience a series of events through a journey. The players need to select their actions to the events, and their choices may apply negative or positive influences on their avatars' attributes. The players need to improve their avatars' attributes to win the battles. After the game is over, the player can submit his avatar data to the blockchain, which enables the avatar to be the part of the story in future journeys. In Last Trip, the developer only provides the basic framework of the story, while the participating players continuously contribute their content to the game itself.

%is a visual novel game based on medieval background. The developer only provides the basic framework of the story, the rest parts are determined by players. The blockchain is combined with the game to provide player created stories. These stories can be Easter eggs, branch missions or puzzles. When a player is dead, his data can be uploaded to the blockchain. This will make him become a Non-Player Character (NPC) to guide the next player. He can choose to be a kind helper or a  trickster depending on his decisions during the game.

\textbf{Adam's Venture\footnote{http://aa.matrixDApp.com}; Ethereum \& TRON \& Nebulas; 3/2019}: Adam's Venture is a multi-player Dungeons and Dragons (D\&D) novel-adventure game. In this game, the players will experience three modes: Battle of Adventure, Battle of Dark Lord, and Battle of Blood Moon. In the Battle of Adventure mode, massive players will create avatars to adventure in the Dungeon, conquer demons to empower their avatars on their own. All growth in these avatars will eventually be accumulated into Adam, a shared character stored in the blockchain. In the battle of Dark Lord, the player will use the shared Adam and one random character summoned from the Last Trip to combat the Dark Lord, the evil king created by the system. Each 30 times Dark Lord is defeated in every chain, the battle of Blood Moon can be triggered. The player will use the Adam in their corresponding chain to conquer the Adams in other chains. According to the recent publication \cite{CaiW2019}, the Adam's Venture and Last Trip are using the same data set for avatars, which means the players are generating contents for both games at the same time.

\textbf{Crypto Space Commander\footnote{https://www.csc-game.com/}; Ethereum; 4/2019}: Crypto Space Commander (CSC) is a sandbox space MMO, that operates a player-owned economy. Players can travel to different star systems, mine planets for resources, craft items and ships to sell, battle with pirates or other players while commanding their unique spaceships. CSC utilizes the Ethereum blockchain to secure game assets, enforce P2P contract actions, and run the game economy. Almost everything can be crafted in the CSC universe. In the beginning, players will be limited to crafting modules, components, and spaceships. Future updates will include the ability for players to craft space stations, resource harvesters, and more. The CSC crafting system encourages players to make unique crafts by giving out rewards. By focusing on their skills and efforts, players can create truly valuable in-game items.

\subsection{Summary: Game Categories}

\begin{figure}[htp]
\centering
\includegraphics[width=8.763cm]{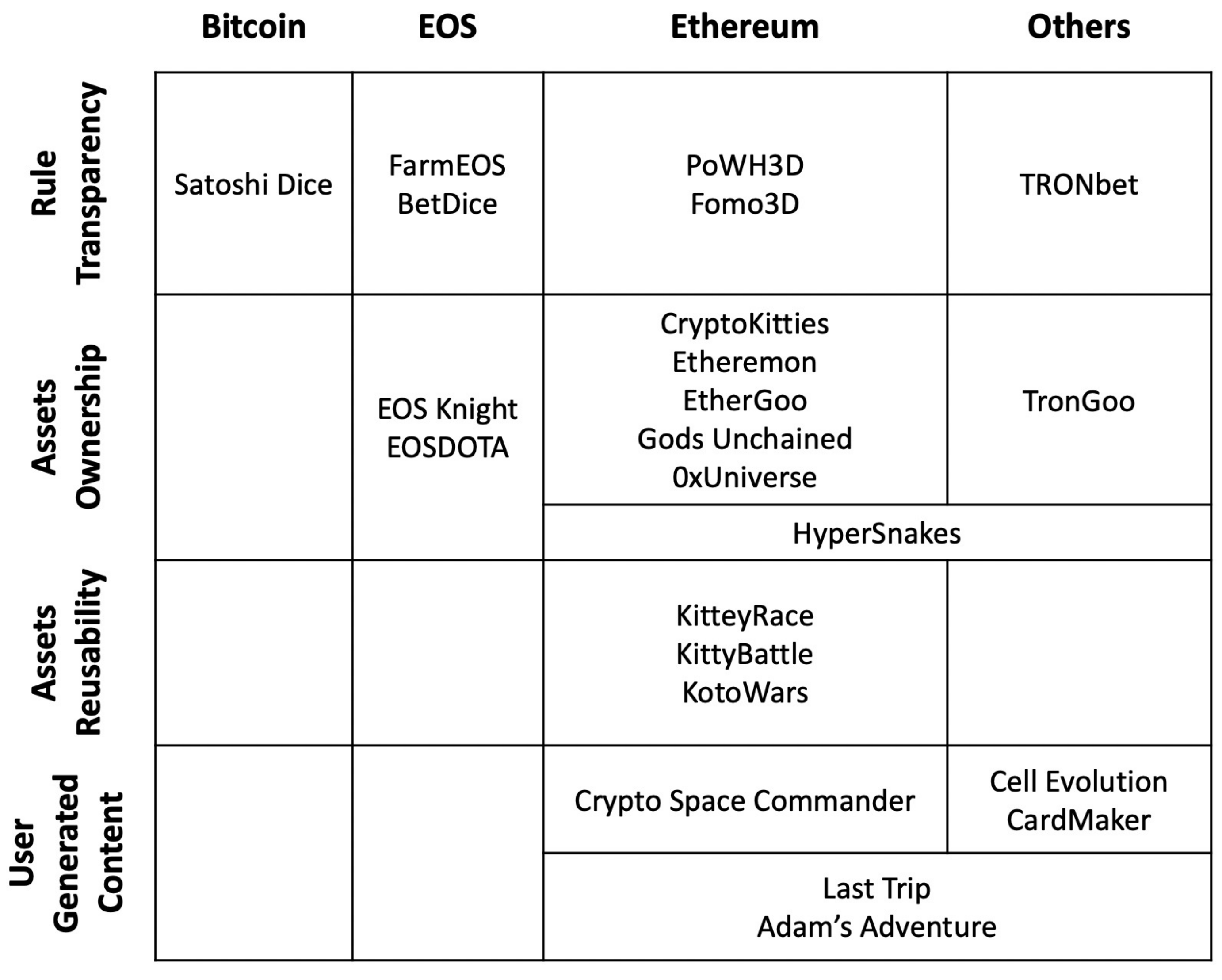}
\caption{Blockchain Game Categories}
\label{Game Categories}
\end{figure}

Fig. \ref{Game Categories} summarized the classifications for the selected blockchain games. Apparently, the technical limitations of blockchain platforms may restrict the blockchain game design. For example, Bitcoin is designed as a decentralized ledger, so that it might not suitable for advanced blockchain applications. On the other hand, Ethereum and EOS provide strong support for smart contract execution, which allows them to implement novel blockchain games. We believe that along with the progress of the technologies, new blockchain platforms in  game vertical will be released to the market to further boost this promising industry.

% Table \ref{game_categories} summarized the above blockchain games by their categories and platforms.

% Actually, the first two characteristics of blockchain provided to games are fundamental requirements: every decentralized application should be transparent and keep ensure users' property rights. We expect more excellent games that fully explore blockchain's advantages.

\section{Data Analysis}\label{sec:analysis}

To analyze the market trend of blockchain games, we collected data up to 18/11/2018 of all 1,323 DApps running on Ethereum, and top 30 EOS DApps which contributed the majority of active users and transaction volume from DAppRadar\footnote{https://DAppradar.com/}. These DApps are divided into three categories: \textbf{1) Games}: all kinds of blockchain games including \textit{Gambling Games} (e.g. online casinos), \textit{Collectible Games}, and \textit{Traditional Games} (e.g. RPG, MMO, TCG, etc). \textbf{2) Trade \& Investment:} including DApps relate to currency and business, such as exchanges, auctions, marketplaces, and Ponzi Schemes. \textbf{3) Others:} including social networking software, non-profit apps, protocols and so on. Two measure criteria are used in the analysis: \textbf{Weekly/Daily Active Users} to measure the popularity of certain categories of DApps. More active users mean the DApp is more popular. \textbf{Weekly/Daily Transaction Volume}  to describe how much currency is transacted through this category of DApps. It reflects the players' willingness to pay.

\subsection{DApps on Ethereum \& EOS}

\subsubsection{DApps on Ethereum}

\begin{figure}[htp]
\centering
\includegraphics[width=8.5cm]{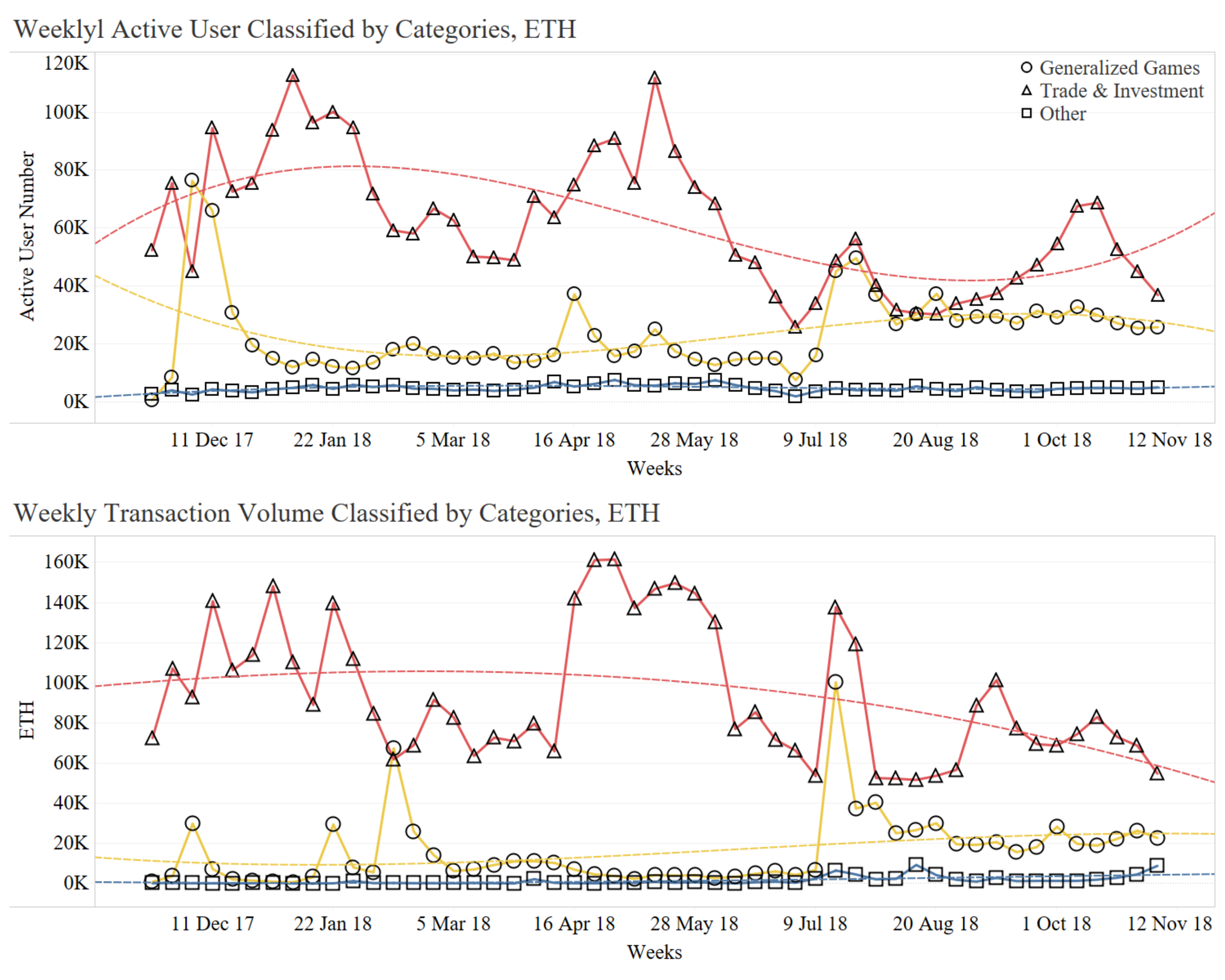}
\caption{DApps on Ethereum}
\label{DApps on Ethereum}
\end{figure}

The top sub-figure in Fig. \ref{DApps on Ethereum} shows that ``Trade \& Investment'' DApps keep having most active users, while its curve fluctuated sharply. ``Games'' had a rapid rise in 11/2017, when the famous game CryptoKitties was released. In a short period of time, players rushed into CryptoKitties for novelty, but soon, a large part of them found it did not meet their expectations and left. Hence, there is a steep decline after the peak. Since then, the number of weekly active users of games on Ethereum has been fluctuating between 10k to 50k. ``Others'' category stay low because of the small quantity of DApps in it. The bottom sub-figure in Fig. \ref{DApps on Ethereum} shows that ``Trade \& Investment'' has dominating transaction volume, but from 9/2018, it has a trend of falling. For the ``Games'', it has several rapid rises and falls. When CryptoKitties released in 11/2017, the curve did not have an enormous ascent, which implies a large group of CryptoKitties players didn't have the will of spending money, or they only spent little ETH on this game. In conclusion, the transaction volume of the ``Games'' category stays in a low condition. The fluctuations didn't significantly bring holistic growth.

\subsubsection{DApps on EOS}

\begin{figure}[htp]
\centering
\includegraphics[width=8.5cm]{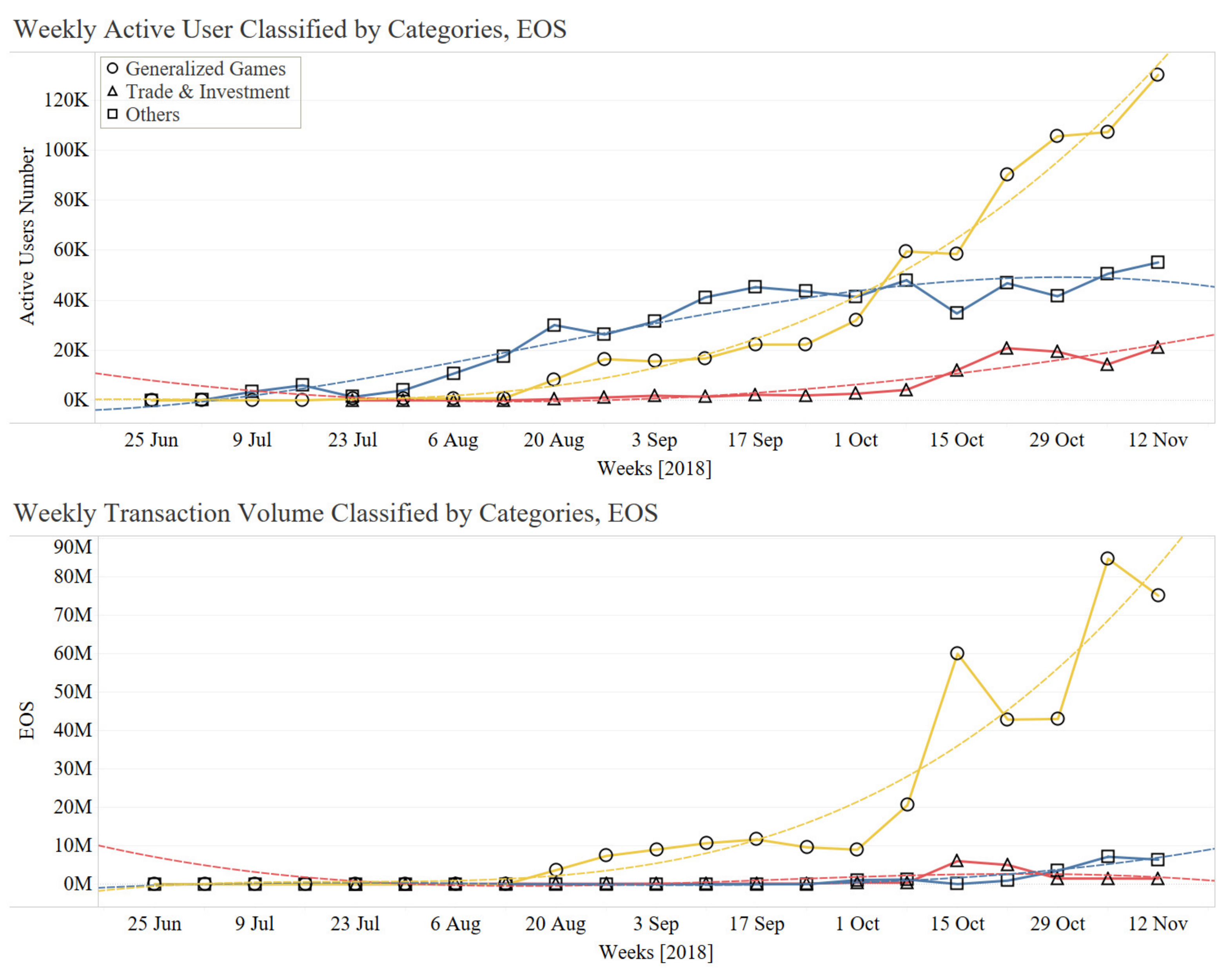}
\caption{DApps on EOS}
\label{DApps on EOS}
\end{figure}

The top sub-figure in Fig. \ref{DApps on EOS} shows that the weekly active DApps user of all categories in EOS tend to go up, especially for the ``Games'', which has gone through an exponential increase from 13/8/2018 this year and exceeded ``Others'' after 8/10/2018. Meanwhile, The bottom sub-figure in Fig. \ref{DApps on EOS} shows the weekly transaction volume of ``Games'' grew rapidly, while the other categories stayed low. Gambling games contributed a lot to this result because casinos want the real-time transaction to speed up the round of betting, and EOS makes it possible.

\subsection{Games on Ethereum \& EOS}

\subsubsection{Active User}

\begin{figure}[htp]
\centering
\includegraphics[width=8.763cm]{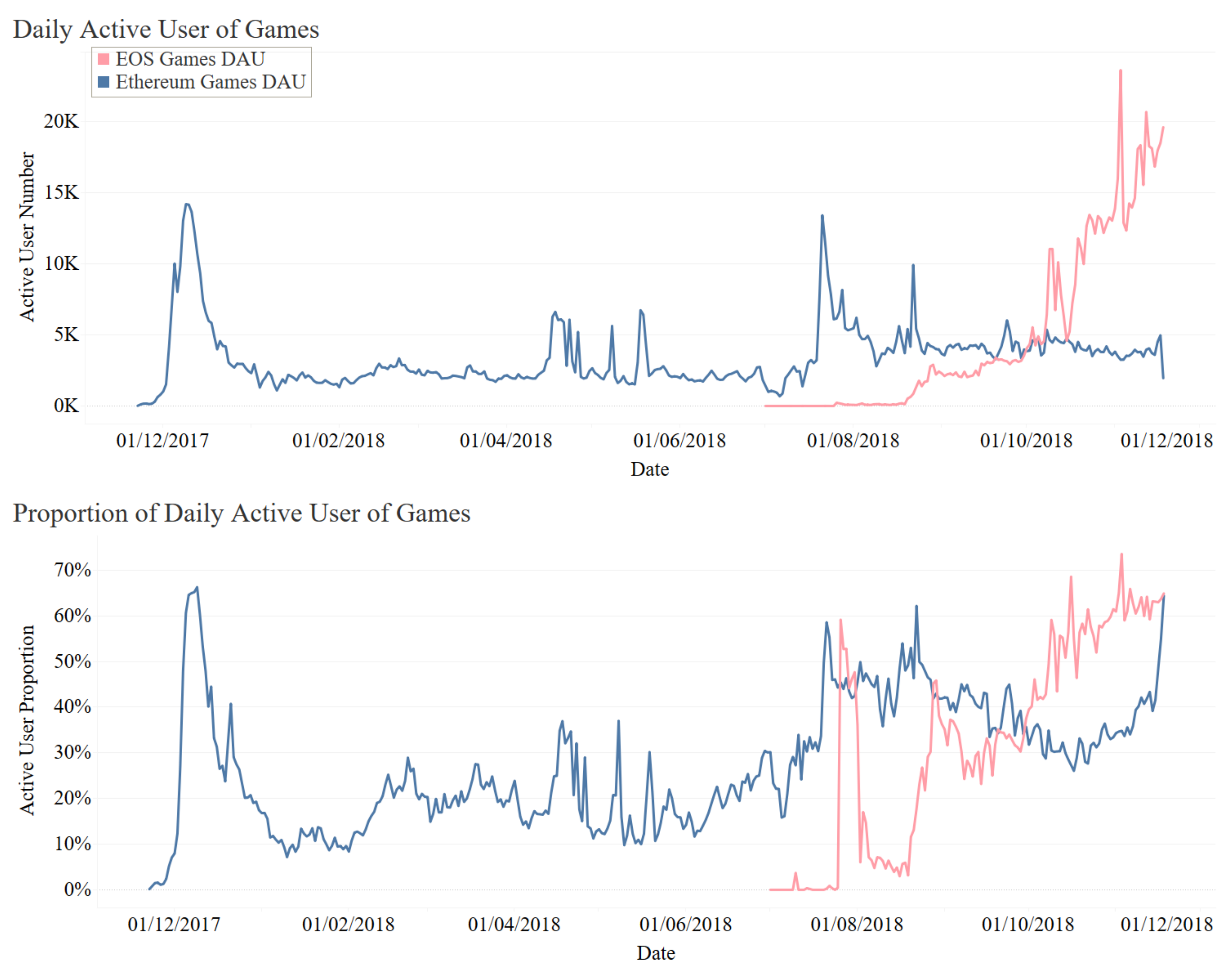}
\caption{Daily Active User of Blockchain Games}
\label{DAU of Games}
\end{figure}

Fig.\ref{DAU of Games} depicts daily active users (DAU) of Ethereum and EOS blockchain games from 11/2017 to 11/2018.

From the top sub-figure, we can see that the Ethereum's DAU once reached its peak of 14,211 on 9/12/2017. After 1/10/2018, the DAU of EOS tended to exceed Ethereum's. EOS curve once reached its local maximum on 3/11/2018, 23651 active users, in one day, and still has a rising trend. In contrast, the Ethereum curve continues to stay low. This may because of the faster transaction and free gas on EOS, which make EOS a more suitable blockchain platform for games to run on. The bottom sub-figure shows the proportion of daily active users of games among all DApps. It is obvious that more than 60\% of DApp users nowadays are game players, which implies the importance of games in the blockchain ecosystem.

% The heat of the first generation blockchain games on Ethereum keeps decaying. Players gradually find these roughly made games boring compared to console games or the Triple-As. Players and developers tend to make money from games rather than enjoying the game-play or making artworks.

\subsubsection{Daily Transaction Volume}

\begin{figure}[htp]
\centering
\includegraphics[width=8.763cm]{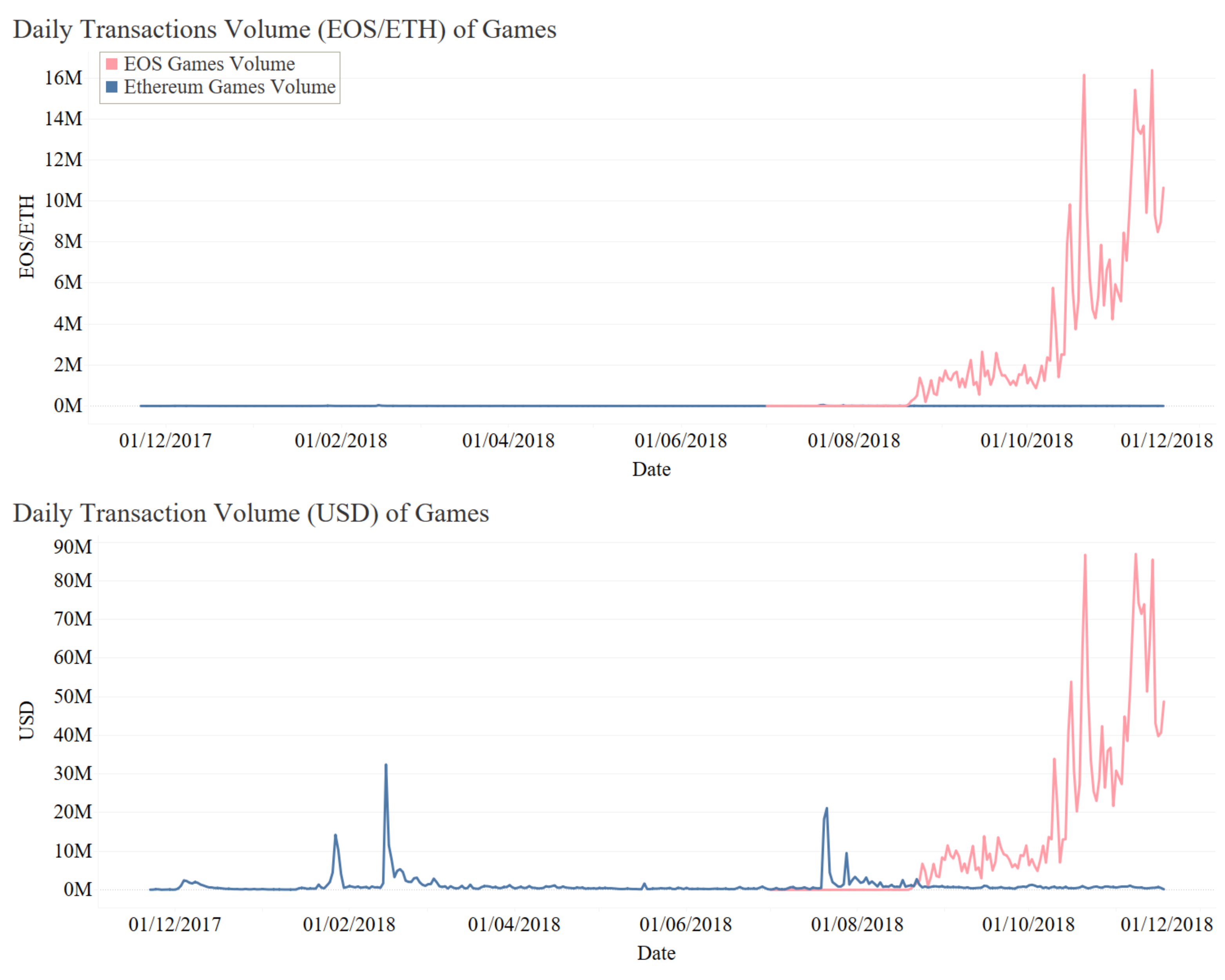}
\caption{Daily Transaction Volume of Blockchain Games}
\label{Transaction of Games}
\end{figure}

Fig. \ref{Transaction of Games} shows that EOS has an absolute advantage on transaction volume. Apart from inflation: EOS is cheaper than ETH, one of the important reasons is that zero gas fee attracted many gambling games and a great number of gamblers, which greatly increases EOS's daily transaction volume. Fluctuations that almost vertical could be pay-out of enormous prizes or bets pouring into the jackpot.

\subsection{Detailed Classified Games on Ethereum}

\begin{figure}[htp]
\centering
\includegraphics[width=8.763cm]{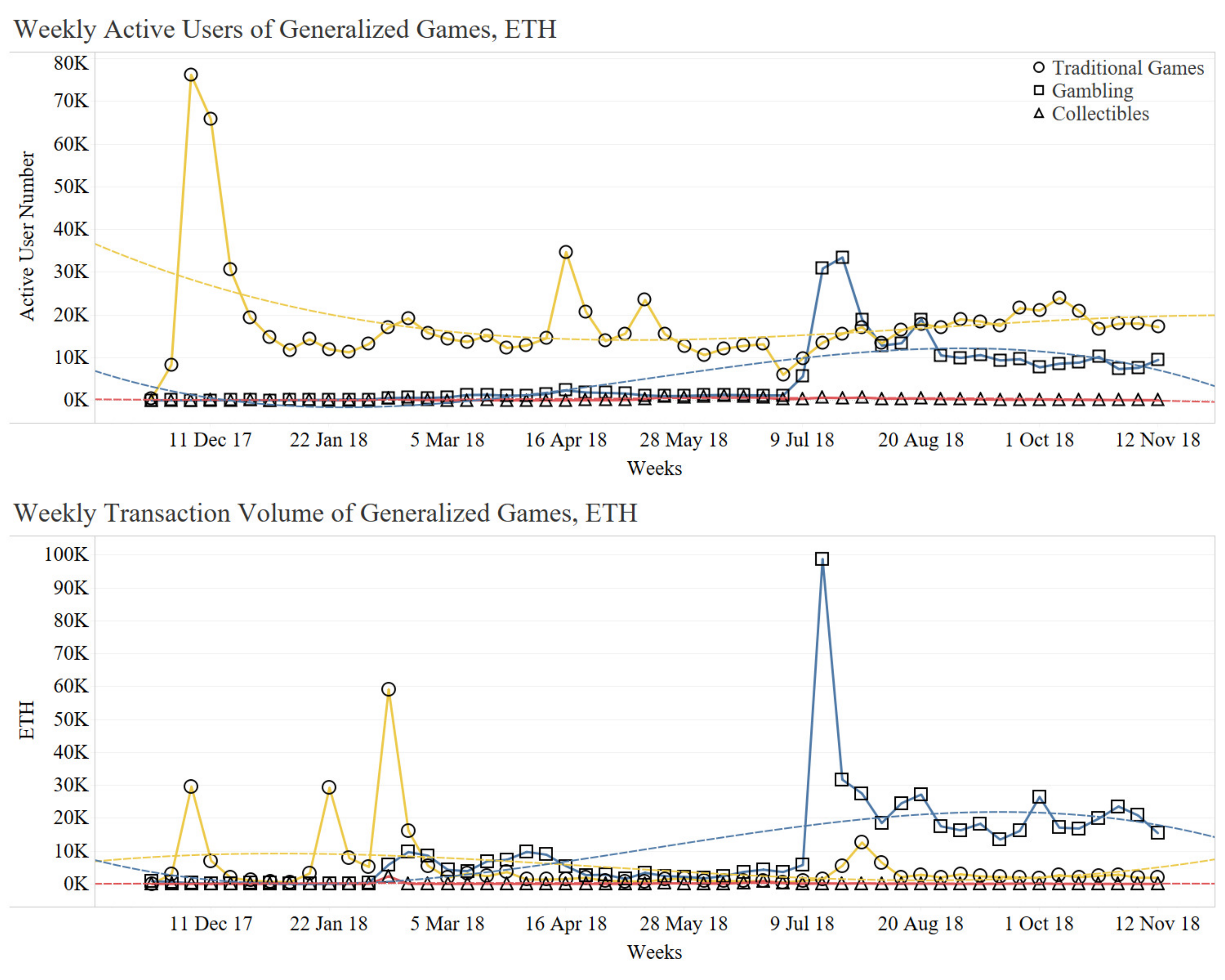}
\caption{Games on Ethereum}
\label{Games on Ethereum}
\end{figure}

Fig. \ref{Games on Ethereum} shows that the contribution of different categories in games that supported the ``Games'' curve in the above section. ``Traditional Games'' had most active users on average, while the transaction wasn't staying in high volume. ``Gambling'' games had significant ascent in both daily active user and transaction volume since 9/7/2018, a leading DApp appeared. ``Collectible'' games stayed low at all time, can be explained by they were bored and not suitable for investment.

\subsection{Detailed Classified Games on EOS}
\begin{figure}[htp]
\centering
\includegraphics[width=8.763cm]{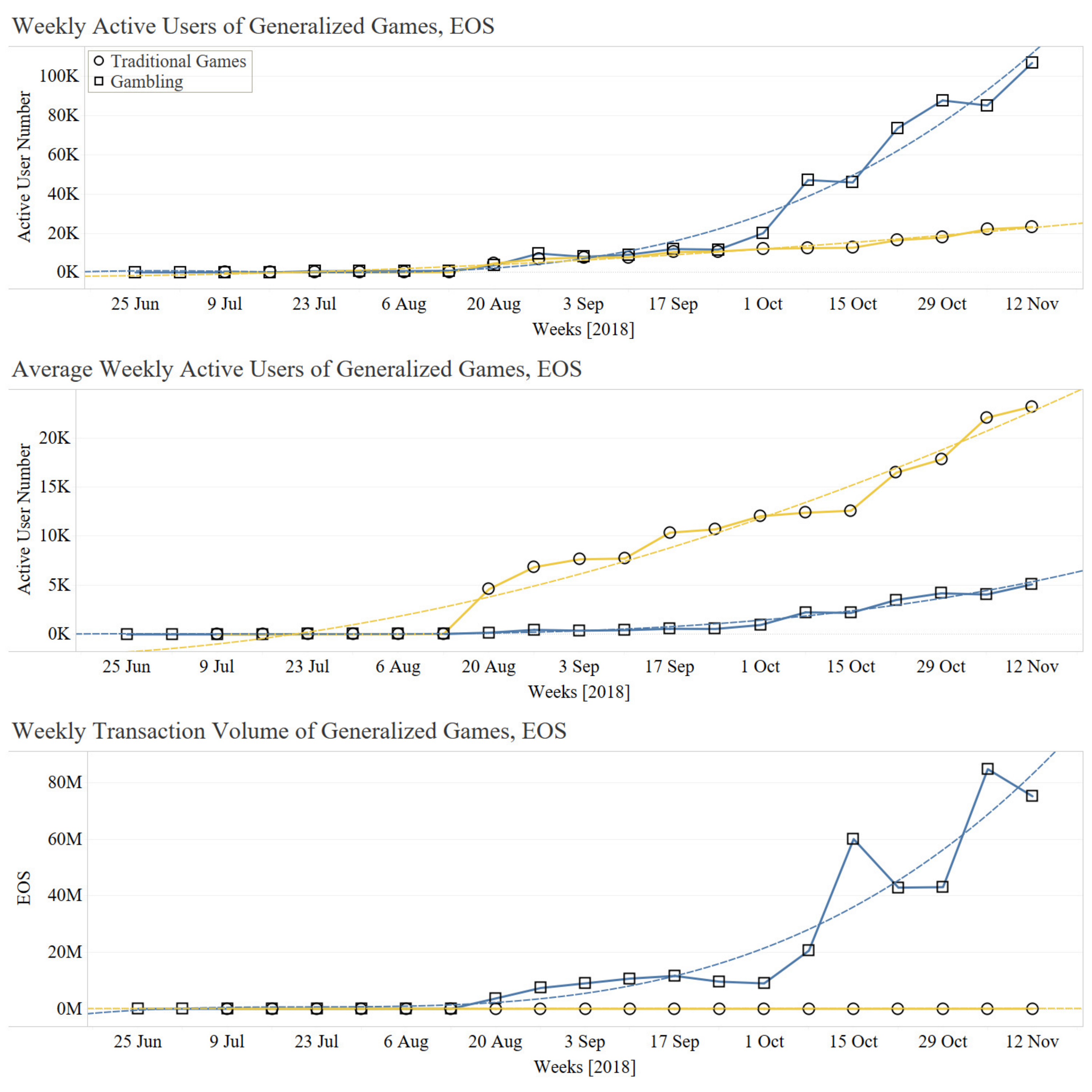}
\caption{Games on EOS}
\label{Games on EOS}
\end{figure}

The top sub-figure in Fig. \ref{Games on EOS} shows the weekly active user of top 30 EOS games. Obviously, ``Gambling'' games player are much more than ``Traditional'' games player. Considering that there is a huge disparity between the number of DApps in two categories, the average weekly active user number of each category was calculated. The middle sub-figure of Fig. \ref{Games on EOS} shows that the active user number of ``Traditional'' Games far exceeds ``Gambling'' games, which is because there is only one game, EOS Knight\footnote{https://eosknights.io/}, in the top 30 DApps. It can be deduced that if there are going to have more popular games like EOS Knight appear on EOS, the ``Traditional'' Games will have much more active users. The bottom sub-figure in Fig. \ref{Games on EOS} shows the weekly transaction Volume of Games on EOS. We can see that ``Gambling'' games contributed the majority in total volume. ``Traditional games'' transaction was much lower, indicating that players did not need to spend much money on it. These games attracted real players and got rid of speculators looking for profit.

\subsection{Analysis Summary}
From macro-perspective, each outstanding milestone DApp can cause explosive user growth. However, the lack of sustainability and consumer stickiness become a problem. This sudden increment will go through a steep decline in a short time. But after the reduction, the number of active users can stabilize at a higher point compared to the value before the surge. Blockchain game needs more leading products like CryptoKitties or EOS Knight to drive the growth of the whole industry and attract more public attention.

From micro-perspective, blockchain games should return to games' original intention: entertaining the players. However, most blockchain games on the market now are still lack of playability. This phenomenon can be explained from a number of perspectives. First, the limitation of blockchain's functionality, which directly makes blockchain games lose the ability to compete with fully functional traditional online games. Second, over-emphasized financial characteristics. Developers focus too much on implementing trading systems or issuing tokens with blockchain. Seldom of them take blockchain as an adjunct of games, or turn it into part of the gameplay. Third, the blockchain game market is still in its infancy. There has been no fierce market competition so far as the traditional game market. Game engines like Unity and Cocos are in an initial stage of integrating blockchain game development packages. Till the end of 2018, no famous studio or publisher has officially announced its entry into the public chain game market or starting their own platform. These imply that blockchain games have huge potential in the future.

% \section{Challenges and Opportunities in Game Research}\label{sec:challenges}

% \subsection{Game Technology}

% \subsubsection{Blockchain for Games}

% new consensus models designed for games (Proof of games, etc.)

% \subsection{Game Mechanics}

% last trip, adam's adventure \cite{CaiW2019} etc.

% \subsection{Narrative}

% \subsubsection{Game}

% \subsubsection{Cost}

% \subsection{Potential Solutions}
% \subsubsection{Novel Consensus Models}
% \subsubsection{Payment Channels}
% \subsection{Security Issues of DApps}
% Smart contracts, executable protocols, are first mentioned by Nick Szabo in 1997\footnote{https://ojphi.org/ojs/index.php/fm/article/view/548/469}. Nowadays, blockchains provide immutable, irreversible, and traceable systems that allow smart contracts to run on, and greatly increased their credibility. However, smart contracts are still in the early stage. Audition tools and bug fixing methods are imperfect, since these codes hard to be modified once deployed. According to a
% \subsection{Common Vulnerabilities}

\section{Conclusion and Vision}\label{sec:conclusion}

Blockchain games benefit from the features of DApps: non-fungible tokens and system transparency. Though these games are still in their preliminary stage, the relationship between players and game companies have been completely changed through such a new concept. Blockchain games may introduce new research directions in the following fields: 1) Game Technologies: new consensus models or alternative technologies for blockchain games should be investigated, on the purpose of improving players' gaming experience with blockchains; 2) Game Mechanics: new mechanics from the idea of blockchain should be developed to enrich the gaming contents. For example, recent work \cite{CaiW2019} has investigated the possibility of asynchronous multi-player gaming experiences across different games; 3) Narrative: the blockchain has introduced a future of immutable data for the virtual world, which may contribute the imagination of game narratives, especially from the perspective of the immortality of life and spirit. 4) Gamified Crowdsourcing: crowdsourcing is an emerging technology that helps to harvest workforce from crowd users. Traditional crowdsourcing platforms are facing challenges like the malicious user, privacy leaks, high costs and insufficient participants. Blockchain games provide a perfect solution to these issues. First, blockchain \cite{Deng2018} enables anonymous, transparent and auditable incentive calculation and distribution, while gamified crowdsourcing \cite{gamifiedcrowdsourcing} may better recruit those users who are insensitive to monetary incentives. Overall, we believe that blockchain will become a disruptive technology for the game industry. The blockchain-empowered virtual worlds in games will eventually lead to real Utopias.

% \section{Challenges and Opportunities in Game Research}\label{sec:challenges}

% \subsection{Game Technology}

% \subsubsection{Blockchain for Games}

% new consensus models designed for games (Proof of games, etc.)

% \subsection{Game Mechanics}

% last trip, adam's adventure \cite{CaiW2019} etc.

% \subsection{Narrative}

\ifCLASSOPTIONcompsoc
  % \section*{Acknowledgments}
\else
  % \section*{Acknowledgment}
\fi


% \ifCLASSOPTIONcaptionsoff
%   \newpage
% \fi

% The ``Triggered'' command can be changed if desired:


%\bibliographystyle{IEEEtran}

%\begin{thebibliography}{1}


%\end{thebibliography}

% \bibliographystyle{ACM-Reference-Format}
\bibliographystyle{ieeetr}
\bibliography{library}

% \begin{IEEEbiographynophoto}{Tian Min}

% \begin{IEEEbiographynophoto}{Hanyi Wang}

% \begin{IEEEbiographynophoto}{Yaoze Guo}

% \begin{IEEEbiography}[{\includegraphics[width=1in,height=1.25in,clip,keepaspectratio]{photos/wcai.jpg}}]{Wei Cai} [S'12-M'16]  received the B.Eng. degree
% in software engineering from Xiamen University, China, in 2008, the M.S. degree in electrical engineering and computer science from Seoul National University, South Korea, in 2011, and
% the Ph.D. degree in electrical and computer engineering from The University of British Columbia (UBC), Vancouver, Canada, in 2016. From 2016 to 2018, he was a Post-Doctoral Research Fellow with UBC. He joined the School of Science and
% Engineering, The Chinese University of Hong Kong, Shenzhen, in 2018, where he is currently an Assistant Professor. He has completed visiting research at the National Institute of Informatics, Japan, The Hong Kong Polytechnic University, and Academia Sinica, Taiwan. His recent research interests include software systems, cloud and edge computing, blockchain systems, and video games. He was a recipient of the 2015 Chinese Government Award for the Outstanding Self-Financed Students Abroad, the UBC Doctoral Four-Year-Fellowship from 2011 to 2015, and the Brain Korea 21 Scholarship. He also received the Best Paper Awards from CloudComp2013, CloudCom2014, and SmartComp2014.
% \end{IEEEbiography}

\end{document}